\documentclass{JFM-FLM_Au}

\usepackage{color,bm}
\usepackage{graphicx}
\usepackage{lineno}
\usepackage{soul}

\newcommand{\lr}[1]{\left(#1\right)}
\newcommand{\mean}[1]{\left\langle#1\right\rangle}
\newcommand{\bnab}{\bm \nabla}
\newcommand{\kmeta}{k_{\max}\ell_\nu}

\newcommand{\be}{\begin{equation}}
\newcommand{\ee}{\end{equation}}

\lefttitle{V.J. Valad\~{a}o, F. De Lillo, S. Musacchio and G. Boffetta}
\righttitle{Scaling and Predictability in SQG Turbulence}

\title{Scaling and Predictability in Surface Quasi-Geostrophic Turbulence}

\author{V.J. Valad\~{a}o\aff{1}, F. De Lillo\aff{1}, 
S. Musacchio\aff{1} \and G. Boffetta\aff{1}}

\affiliation{\aff{1}Dipartimento di Fisica and INFN - Universit\`a degli 
Studi di Torino, Via Pietro Giuria, 1, 10125 Torino TO, Italy.}

\corresau{V.J. Valad\~{a}o, \email{victor.dejesusvaladao@unito.it}}

\begin{document}
\maketitle
\begin{abstract} 
  Turbulent flows are strongly chaotic and unpredictable, with a Lyapunov exponent that increases with the Reynolds number. Here, we study the chaoticity of the Surface Quasi-geostrophic system, a two-dimensional model for geophysical flows that displays a direct cascade similar to that of three-dimensional turbulence. Using high-resolution direct numerical simulations, we investigate the dependence of the Lyapunov exponent on the Reynolds number and find an anomalous scaling exponent larger than the one predicted by dimensional arguments. We also study the finite-time fluctuation of the Lyapunov exponent by computing the Cram\'er function associated with its probability distribution. We find that the Cram\'er function attains a self-similar form at large $\text{Re}$.
  
\end{abstract}
\begin{keywords}
    Surface Quasi-geostrophic, Turbulence, Predictability, Direct Numerical Simulations.
\end{keywords}

\section{Introduction}
\label{sec1}

Turbulence is a complex and chaotic phenomenon characterized by a large number
of interacting degrees of freedom organized hierarchically across multiple
scales of motion \citep{frisch1995turbulence}. Determining whether a turbulent flow
remains predictable or, at each scale, retains any degree of
predictability has been a longstanding challenge, tracing back to the
pioneering works by Lorenz, Ruelle, Leith, and Kraichnan 
\citep{lorenz1969predictability,leith1971atmospheric,leith1972predictability,ruelle1979microscopic,deissler1986navier}. 
The nonlinear
amplification of small-scale perturbations led to the formulation of the famous
``butterfly effect'' \citep{lorenz1963deterministic}. When extended to multiscale
systems such as turbulence, these ideas give rise to the concept of the cascade
of errors, in which small-scale perturbations progressively amplify and
propagate to larger scales, gradually spoiling predictability at larger and
larger scales \citep{boffetta1997predictability, rotunno2008generalization,
palmer2024real,boffetta2017chaos}. 

A key mathematical tool for studying predictability is the Lyapunov 
exponent and its finite-time version (FTLE). 
Ruelle predicted that the Lyapunov exponent in turbulence is proportional
to the inverse of the smallest time, the Kolmogorov time, and therefore,
it increases with the flow's Reynolds number ($\text{Re}$) \citep{ruelle1979microscopic}.
Nonetheless, a turbulent flow at large $\text{Re}$ remains predictable
at large scales since the error grows with the characteristic turnover time 
of that scale, which is independent of the Reynolds number. 
This property is familiar in oceanography and atmospheric
flows where the smallest time scale 
can be very small (fraction of seconds) \citep{garratt1994atmospheric} 
while the weather remains predictable for days.

In this work, we consider the surface quasi-geostrophic (SQG) equation,
a model that describes the flow governed by the conservation of buoyancy 
at the surface of a rotating, stratified fluid 
\citep{blumen1978uniform,pierrehumbert1994spectra}. Beyond
its geophysical relevance, the SQG model has gained attention in the 
fluid dynamics community due to its striking similarities to 
three-dimensional (3D) Navier-Stokes (NS) turbulence while keeping some properties 
of two-dimensional (2D) flows. In particular, SQG has two inviscid quadratic 
invariants, similarly to 2D turbulence 
\citep{celani2004active,lapeyre2017surface,valade2024anomalous}
with one of the two, the surface
potential energy, which displays, in the presence of forcing and dissipation,
a direct cascade \`a la Kolmogorov towards the small scales similar to 3D turbulence
\citep{valadao2024nonequilibrium}.
Despite this similarity, previous numerical studies
\citep{pierrehumbert1994spectra,ohkitani1997inviscid,sukhatme2002surface}
reported that the scaling exponent of the spectrum of the surface potential energy
deviates from the Kolmogorov value $-5/3$ predicted by dimensional arguments. 

Based on very high-resolution direct numerical simulations of the 
SQG model, we study the predictability of the direct cascade in relation to the 
scaling properties of the surface buoyancy field.
At large Reynolds numbers, we observe the recovery of the Kolmogorov scaling
in the surface potential energy spectrum. 
Further, we measure the finite-time distribution of the 
Lyapunov exponent as a function of $\text{Re}$ and we find an anomalous scaling law
in which the Lyapunov exponent grows faster than what is predicted on 
dimensional grounds, similar to what was observed in 3D NS turbulence 
\citep{boffetta2017chaos,mohan2017scaling,berera2018chaotic}. Despite this anomaly, we find that the distribution of the 
finite-time Lyapunov exponents follow an almost universal function, 
independent of the Reynolds number of the flow. 

The remainder of this paper is as follows. In section \ref{sec2} we
review the basic definitions of the SQG equation and discuss its statistical
properties in the turbulent regime. Section \ref{sec3} discusses the main 
results obtained in this work through the use of extensive numerical 
simulations covering a large range of Reynolds numbers. We split the results
into two subsections: Section~\ref{sec3a} explores the Reynolds dependence on the
dimensional scaling properties of SQG,
principally the scaling exponent of the surface potential energy spectrum;
Section~\ref{sec3b} addresses the Eulerian predictability and the statistics of finite-time Lyapunov 
exponent (FTLE) as functions of Reynolds number.
In Section \ref{sec4}, we summarize our results, pointing out directions for future research.

\section{The Surface Quasi-geostrophic model}\label{sec2}

The SQG model describes the large-scale dynamics of a rapidly rotating, stably stratified flow using a two-dimensional equation for the surface buoyancy field
$\theta({\bm x}, t)$ \citep{pierrehumbert1994spectra,juckes1994quasigeostrophic, lapeyre2006dynamics, vallis2017atmospheric,lapeyre2017surface,siegelman2022moist}:  
\begin{equation}
\partial_t \theta + {\bf v} \cdot {\bnab} \theta = \nu \nabla^2 \theta - \mu \nabla^{-2}\theta + f \;.
\label{eq1}
\end{equation}

The incompressibility condition $\bnab\cdot{\bf v}=0$ is enforced through 
the stream function $\psi(\bm x,t)$ which defines the velocity field as 
${\bf v}({\bm x}, t) = (-\nabla_y \psi, \nabla_x \psi)$. 
The relation between the buoyancy field and the stream function is
given by $\psi=\nabla^{-1} \theta$ or, in Fourier space, 
$\hat\psi=\hat\theta/k$ (where $k\equiv|\bm k|$) and consequently the velocity 
field can be written in terms of buoyancy field as  
\begin{equation}
\hat{\bf v}({\bm k}) = \left(-\frac{i k_y}{k}, \frac{i k_x}{k}\right) 
\hat{\theta}({\bm k})
\label{eq2}  
\end{equation}  
from which one observes that $\theta$ has the dimension of a velocity. 

In the absence of forcing and dissipation ($f = 0$, $\nu = 0$, $\mu = 0$), the
SQG equation (\ref{eq1}) conserves two quadratic quantities,
the vertically integrated energy (VIE),  
\begin{equation}
V = \frac{1}{2} \langle \psi \theta \rangle \;,
\label{eq3}
\end{equation}  
and the surface potential energy (SPE),  
\begin{equation}
E = \frac{1}{2} \langle \theta^2 \rangle \;,
\label{eq4}
\end{equation}  
where brackets stand for spatial average.

The dissipative and forcing terms in the SQG equation represent the effects of scales not resolved by the model, their specific form is somewhat arbitrary \citep{smith2002turbulent,lapeyre2006dynamics,burgess2015kraichnan}. In (\ref{eq1}) we introduce a diffusivity $\nu$ and a large-scale damping $\mu$ chosen to be active at, respectively, small and large scales only, while the forcing is active on a narrow range of scales around $\ell_f$.

Under these conditions, one expects that a turbulent flow develops with a double cascade phenomenology
\citep{blumen1978uniform, pierrehumbert1994spectra}.
Within this scenario, SPE is primarily transferred from the forcing scale 
to smaller ones ($\ell < \ell_f$), producing the direct cascade which 
is eventually dissipated by viscosity at the diffusive scale $\ell_\nu$. 
Meanwhile, VIE undergoes an inverse cascade, transferring energy to scales 
larger than the forcing ($\ell > \ell_f$) until it is dissipated at 
the friction scale $\ell_\mu$.
In the statistically stationary state, the SPE and VIE balances are given by  
\begin{equation}
\varepsilon_I = \varepsilon_\nu + \varepsilon_\mu \;,
\label{eq5}
\end{equation}  
\begin{equation}
\eta_I = \eta_\nu + \eta_\mu \;,
\label{eq6}
\end{equation}  
where $\varepsilon_I = \mean{\theta f}$ and $\eta_I=\mean{\psi f}$ are 
the input rates, 
$\varepsilon_\nu = \nu \mean{|{\bnab} \theta|^2}$ 
and $\eta_\nu=\nu\mean{\bnab\psi\cdot\bnab\theta}$ the small-scale
dissipation rates while $\varepsilon_\mu = \mu \mean{\theta \nabla^{-2} \theta}$ 
and $\eta_\mu=\mu\mean{\psi \nabla^{-2} \theta}$ are the large-scale dissipation 
rates of SPE and VIE, respectively. 

Under the assumptions of statistical homogeneity and isotropy, \cite{blumen1978uniform} first predicted the power-law behavior of the energy
spectrum $E(k)\equiv \langle|\hat\theta({\bm k})|^2\rangle/2$ for sufficiently 
large scale
separations $\ell_\nu\ll\ell_f\ll\ell_\mu$. In this case, the spectral energy
densities follow  
\begin{equation}
E(k) \simeq \eta_\mu^{2/3} k^{-1}, \quad 1/\ell_\mu \ll k \ll 1/\ell_f
\label{eq7}
\end{equation}
\begin{equation}
E(k) \simeq \varepsilon_\nu^{2/3} k^{-5/3}, \quad 1/\ell_f \ll k \ll 1/\ell_\nu
\label{eq8}
\end{equation}
and the diffusive and friction scales are determined on dimensional grounds as
\begin{equation}
\ell_\nu \equiv \left(\frac{\nu^3}{\varepsilon_\nu} \right)^{1/4}, \quad \ell_\mu \equiv \left(\frac{\eta_\mu}{\mu^3} \right)^{1/9}.
\label{eq9}
\end{equation}  
The ratio between the dissipative and forcing scales, i.e., the extension
of the inertial range, defines the Reynolds numbers associated with the 
flow. For the direct cascade of SPE, in analogy to 3D NS turbulence,
we define the Reynolds number as 
\begin{equation}
\text{Re} \equiv \frac{\varepsilon_I^{1/3} \ell_f^{4/3}}{\nu} \simeq 
\left(\frac{\ell_f}{\ell_\nu} \right)^{4/3}
\label{eq10}
\end{equation}
Note that in (\ref{eq10}) $\text{Re}$ is based on 
$\varepsilon_I$ and it is therefore defined a priori. 
Alternatively, we could use the VIE dissipation $\varepsilon_{\nu}$
resulting in a slightly smaller value of $\text{Re}$, as discussed below. 

The scaling laws in (\ref{eq7}-\ref{eq8}) can also be obtained from the analogous for SQG of the exact four-fifths law of turbulence \citep{valadao2024nonequilibrium,valade2025surface} and by assuming self-similarity of the statistics \citep{frisch1995turbulence}. Following this approach, one expects velocity and surface buoyancy to have the same scaling exponent $1/3$, which corresponds to that of the velocity field in 3D NS. This observation allows us to adapt to SQG the dimensional arguments developed by Ruelle for the predictability of 3D NS turbulence \citep{ruelle1979microscopic}. According to the latter, the Lyapunov exponent $\lambda$ characterizing the exponential growth of an infinitesimal perturbation of a solution of \eqref{eq1} should be proportional to the inverse of the smallest dynamical time, i.e., the Kolmogorov time $\tau_{\nu}\equiv(\nu/\varepsilon_{\nu})^{1/2}$ 
\begin{equation}
\lambda \simeq \frac{1}{\tau_\nu}\simeq \frac{1}{\tau_f} \text{Re}^{1/2} \ .\
\label{eq21}
\end{equation}
In the following Section, we numerically investigate the scaling properties of the surface buoyancy field and the prediction \eqref{eq21}.

\section{Numerical simulations and results}\label{sec3}

We explore the statistical properties and the Eulerian predictability 
of the direct cascade in SQG at different Reynolds numbers 
by numerically integrating \eqref{eq1} at high resolution with a 
pseudo-spectral, GPU-accelerated code. 
Simulations are performed in a square domain of size $L_x = L_y = 2\pi$ 
with periodic boundary conditions, using a regular grid with resolution 
$N\times N$ ranging from $N = 1024$ to $N=16384$. 
Simulations cover more than two decades in the diffusion coefficient,
corresponding to a Reynolds number \eqref{eq10} which varies from 
$\text{Re}=600$ to $\text{Re}=159000$, while the large
scale dissipation coefficient $\mu$ is fixed. 
For all runs, the system is driven by a constant-amplitude forcing with random phases, active
within a narrow circular shell in wavevector space centered on $k_f=3.5$ and
with a small width $\Delta k=0.5$. 
This forcing provides constant SPE and VIE injection rates 
$\varepsilon_I$ and $\eta_I$ respectively with $\varepsilon_I \approx \eta_I k_f$ 
since $\Delta k \ll k_f$. Specific details
on the GPU code performances can be found in \citep{valadao2025spectrum}.

The most relevant parameters on the simulations are listed in Table~\ref{tab1}. All the simulations are performed in statistically stationary states, including the subset of simulations for computing the Lyapunov exponent. We also performed a careful study on the sensitivity of the following results to the maximum resolved wavenumber $\kmeta$ by increasing resolution at fixed $\text{Re}$. We found independence of the results on the resolution for $\kmeta\gtrsim1.5$.
\begin{table}
\begin{center}
\begin{tabular}{ccccccc||ccccccc}
 Run & $N$ & $\text{Re}$ & $\kmeta$ & $\tau_\nu$ & $T_{tot}/\tau_f$ &&& Run & $N$ & $\text{Re} $ & $\kmeta$ & $\tau_\nu$ & $T_{tot}/\tau_f$ \\
 
 $A_1$ & 1024 &  $642$ & $5.5$ & $0.0260$ &    $-$ &&&  $C_2$ & 4096  &  $10600$ & $2.9$ & $0.0075$ & $1520$ \\
 $A_2$ & 1024 &  $794$ & $4.7$ & $0.0240$ &    $-$ &&&  $C_3$ & 4096  &  $15900$ & $2.1$ & $0.0062$ & $1520$ \\
 $A_3$ & 1024 & $1060$ & $3.8$ & $0.0210$ &    $-$ &&&  $C_4$ & 4096  &  $21200$ & $1.7$ & $0.0054$ & $1520$ \\
 $A_4$ & 1024 & $1590$ & $2.9$ & $0.0180$ &    $-$ &&&  $C_5$ & 4096  &  $25400$ & $1.5$ & $0.0049$ & $1520$ \\
 $B_1$ & 2048 & $2120$ & $4.7$ & $0.0160$ &    $-$ &&&  $D_1$ & 8192  &  $31800$ & $2.5$ & $0.0044$ & $9\times287$ \\
 $B_2$ & 2048 & $3180$ & $3.5$ & $0.0130$ & $2270$ &&&  $D_2$ & 8192  &  $42400$ & $2.1$ & $0.0039$ & $9\times287$ \\
 $B_3$ & 2048 & $6350$ & $2.1$ & $0.0099$ &    $-$ &&&  $D_3$ & 8192  &  $63500$ & $1.5$ & $0.0031$ & $9\times287$ \\
 $B_4$ & 2048 & $7940$ & $1.8$ & $0.0086$ & $2270$ &&&  $E_1$ & 16384 &  $90800$ & $2.3$ & $0.0026$ & $-$ \\
 $C_1$ & 4096 & $3970$ & $5.9$ & $0.0120$ &    $-$ &&&  $E_2$ & 16384 & $159000$ & $1.6$ & $0.0020$ & $-$ \\
\end{tabular}
    \caption{Relevant parameters of the simulation:
    Reynolds number, 
    diffusive scale $\ell_\nu=\nu^{3/4}\varepsilon_\nu^{-1/4}$, 
    diffusive time $\tau_\nu=\sqrt{\nu/\varepsilon_\nu}$, 
    total length of the Lyapunov simulations $T_{\text{tot}}$ 
    ($9$ independent realizations of length $287 \tau_{f}$ for the runs $D$).
    Common parameters for all simulations:
    forcing wavenumber $k_f=3.5$ and width $\Delta k_f=0.5$,
    surface potential energy input $\varepsilon_I=24$,
    friction coefficient $\mu=1.0$,
    characteristic time at the forcing scale $\tau_f=\varepsilon_I^{-1/3}\ell_f^{2/3}=0.51$,
    maximum resolved wavenumber $k_{\max}=N/3$ (2/3 dealiasing rule).}
    \label{tab1}
\end{center}
\end{table}

\subsection{Statistics of the direct cascade}\label{sec3a}

For large separations between the forcing and the dissipative scales,
one expects that almost all the SPE is transferred and dissipated at
small scales, while VIE is dissipated at large scales. This is the 
essence of the argument developed by Fjortoft for 2D turbulence 
\citep{fjortoft1953changes} and verified by numerical simulations 
of 2D NS double cascade \citep{boffetta2010evidence}. 
In the present case, using (\ref{eq5}) and (\ref{eq6}) and the scaling 
relation $\eta_{\ell} \simeq \ell \varepsilon_{\ell}$ this argument
gives 
\begin{equation}
\frac{\eta_\nu}{\eta_\mu}=
\lr{\frac{\ell_\mu-\ell_f}{\ell_f-\ell_\nu}}\frac{\ell_\nu}{\ell_f} \simeq 
\text{Re}^{-3/4}
\label{eq11}
\end{equation}
where we have used (\ref{eq10}) to express $\ell_{\nu}/\ell_f$ as a function of $\text{Re}$. 

\begin{figure}[h]
\centering\includegraphics[width=0.75\linewidth]{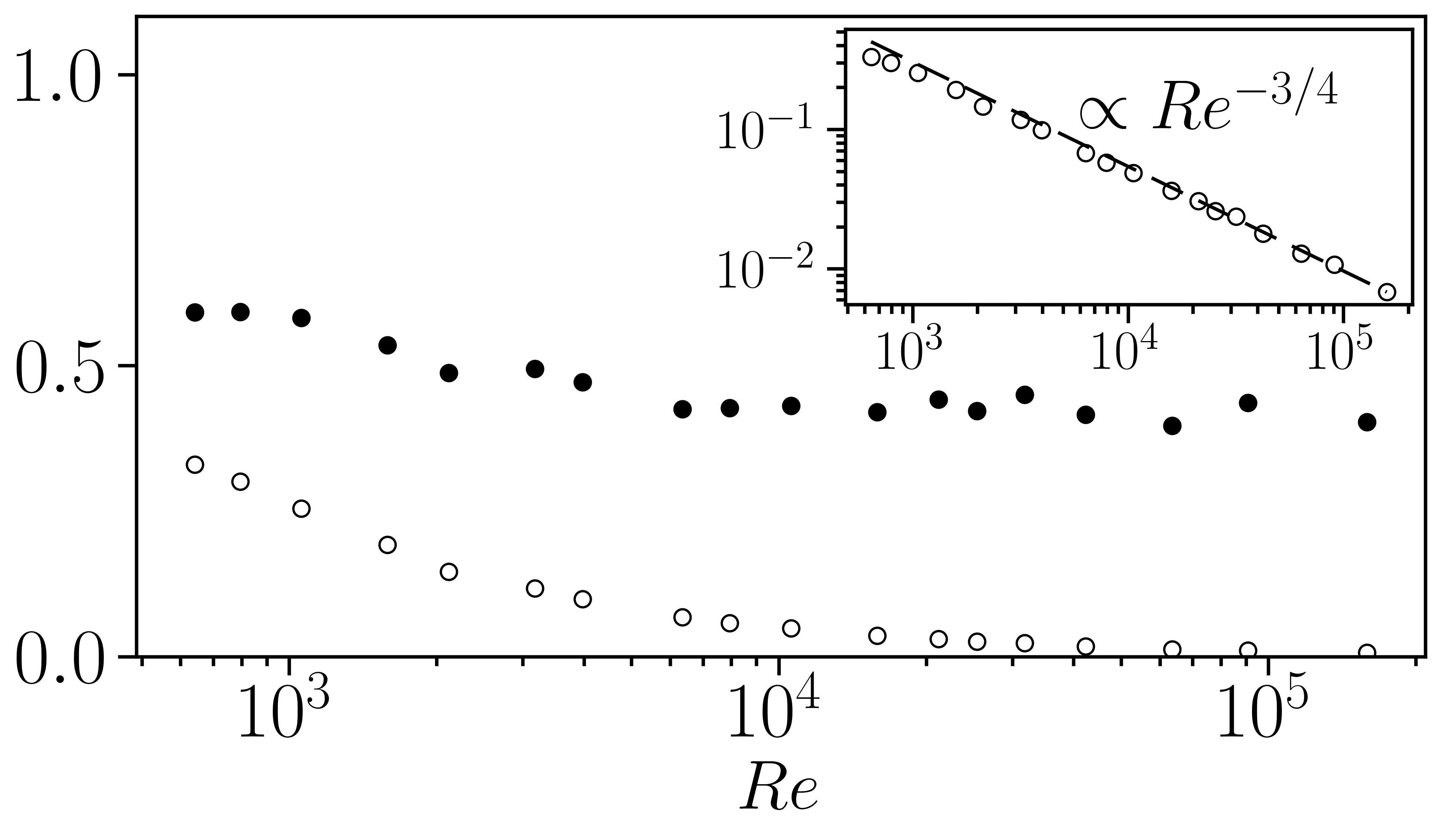}
\caption{Main figure: SPE relative dissipation $\varepsilon_\nu / \varepsilon_I$ (full circles) and
VIE relative dissipation $\eta_\nu / \eta_I$ (open circles), both as functions of $\text{Re}$.
The inset shows $\eta_\nu/\eta_I$ as functions of $\text{Re}$ on a log-log scale.}
\label{fig1}
\end{figure}

Figure~\ref{fig1} shows the fraction of small-scale dissipation of
the inviscid invariants as a function of the Reynolds number of the flow
in statistically stationary conditions. 
Indeed, we find that the small-scale relative dissipation of VIE vanishes
in the limit of large $\text{Re}$ following the prediction \eqref{eq11}. 
On the contrary, since we do not resolve the inverse cascade and
$\ell_{\mu} \sim \ell_f$, there remains a constant large-scale SPE
dissipation $\varepsilon_\mu$ for large $\text{Re}$. 
Therefore, the direct cascade transfers only a fraction of the total injected
energy equivalent to $\varepsilon_\nu\approx0.45\varepsilon_I$ for $\text{Re}\gtrsim10^4$.
\begin{figure}[!ht]
\centering\includegraphics[width=0.75\linewidth]{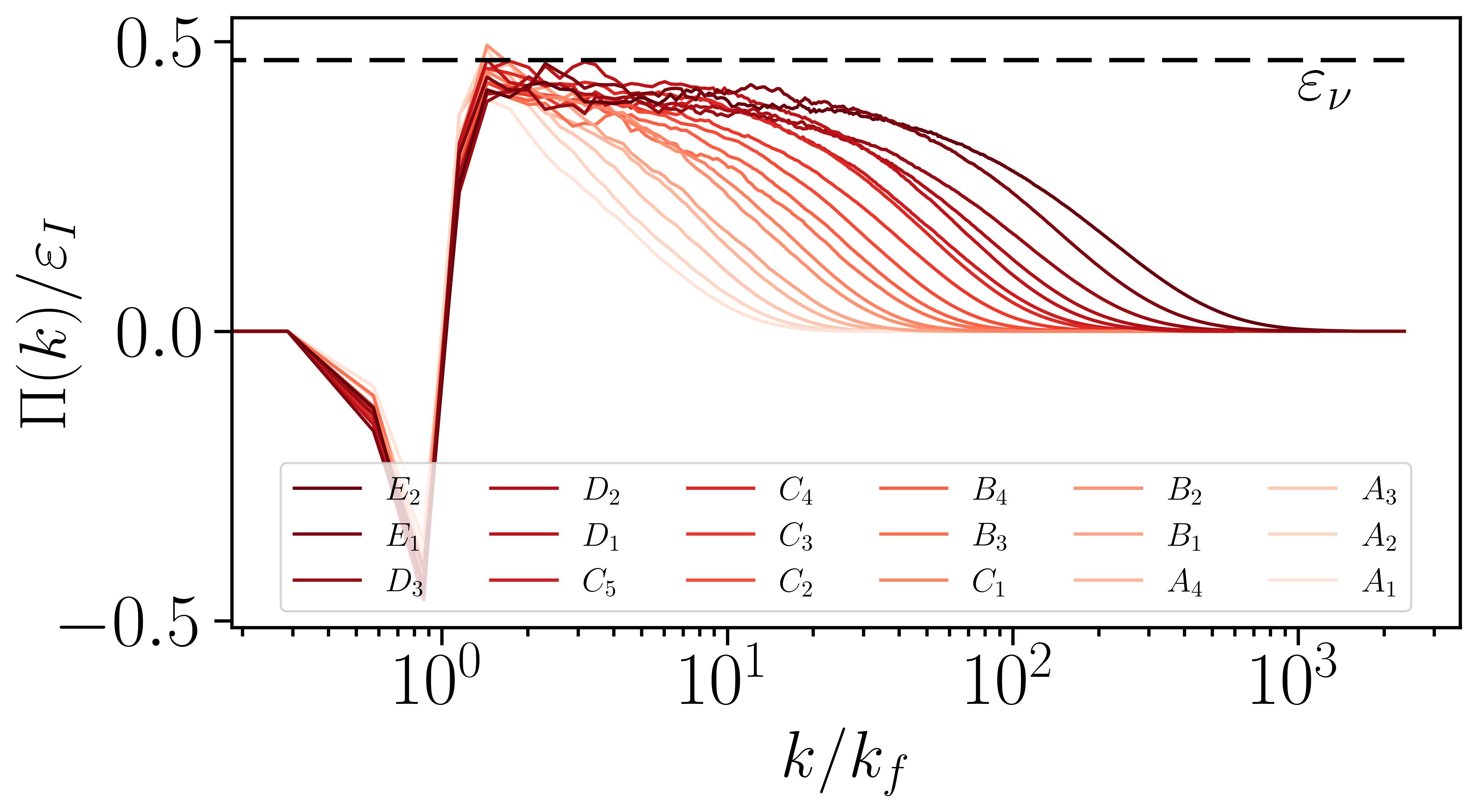}
\caption{Normalized SPE direct cascade fluxes $\Pi(k)$ of all runs on table~\ref{tab1}. The dashed line represents $\varepsilon_\nu / \varepsilon_I=0.45$.}
\label{fig2}
\end{figure}

Stationary fluxes of SPE in Fourier space are presented 
in Fig.~\ref{fig2} for different
Reynolds numbers. At moderate $\text{Re} \lesssim 3000$ the fluxes for $k>k_f$ 
decay quickly as a consequence of the viscous dissipation. For large $\text{Re}$,
however, a plateau of constant flux emerges at a level corresponding to the
viscous dissipation rate $\varepsilon_\nu$. 
We emphasize that SQG turbulence exhibits large
fluctuations in the flux of the direct cascade, as studied in details
in \citep{valadao2024nonequilibrium}.
These fluctuations arise from
the interplay between the accumulation of energy in large-scale structures and
intense dissipative events triggered by the formation of filamentary shocks
that transfer energy from large to small scales over short time intervals.
Thus, very long integrations are necessary to observe the convergence to the
constant flux plateau of Fig.~\ref{fig2}.
\begin{figure}[!ht]
\centering\includegraphics[width=0.75\linewidth]{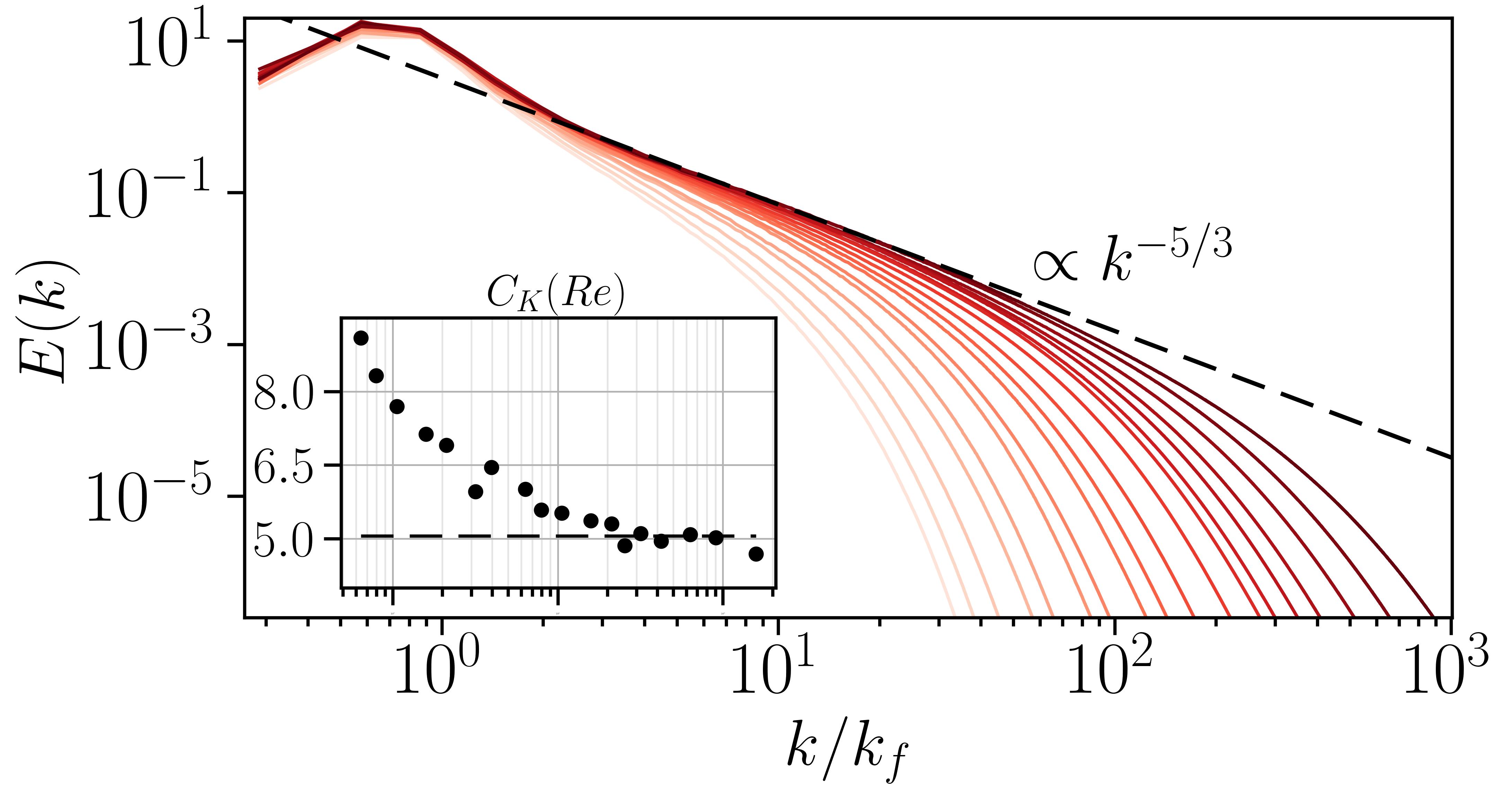}
\caption{Main plot: Time-averaged spectra $E(k)$ for all simulations. Color coding
follows the same as Fig.~\ref{fig2}.
Inset: $C_K$ as functions of the Reynolds number. 
}
\label{fig3}
\end{figure}
 
The time-averaged spectra $E(k)$ of SPE are shown in Fig.~\ref{fig3}
for all the runs in Table~\ref{tab1}.
All spectra exhibit power-law behavior, $E(k) \propto
k^{-\beta}$ in an intermediate range of scales, which becomes wider as 
$\text{Re}$ increases. 
We observe that at moderate $\text{Re}$, when a scaling range is
already clearly observable, the scaling exponent $\beta$ deviates 
significantly from the dimensional prediction $5/3$, a feature already 
reported by previous investigations at comparable Reynolds numbers
\citep{pierrehumbert1994spectra,ohkitani1997inviscid,sukhatme2002surface}.
Nonetheless, we find that when $\text{Re}$ is sufficiently large, $\text{Re} \gtrsim 10^4$,
the scaling of the dimensional prediction (\ref{eq8}) is closely
recovered. 

To quantify this important result, we measured the 
correction $\xi$ to the dimensional scaling exponent by fitting the
intermediate range of the spectra with 
\begin{equation}
E_K(k) = C_K \varepsilon_\nu^{2/3} k^{-5/3} \left(\frac{k}{k_f}\right)^{-\xi}
\label{eq12}
\end{equation}  
where $\xi$ and $C_K$ are the fitting parameters. 
In order to estimate the robustness of the fit, we adopted the following 
procedure: For each run, we fit the data with \eqref{eq12} in a range
of wavenumbers $k \in [k_0, k_1]$ with varying 
$k_0\in[3k_f, 5k_f]$ and $k_1\in[8k_f, 10k_f]$. This produces a set
of parameters $\xi$ and $C_K$ for each run, from which we compute the 
mean using twice the standard deviation as an estimation of the error.

\begin{figure}[!ht]
\centering\includegraphics[width=0.75\linewidth]{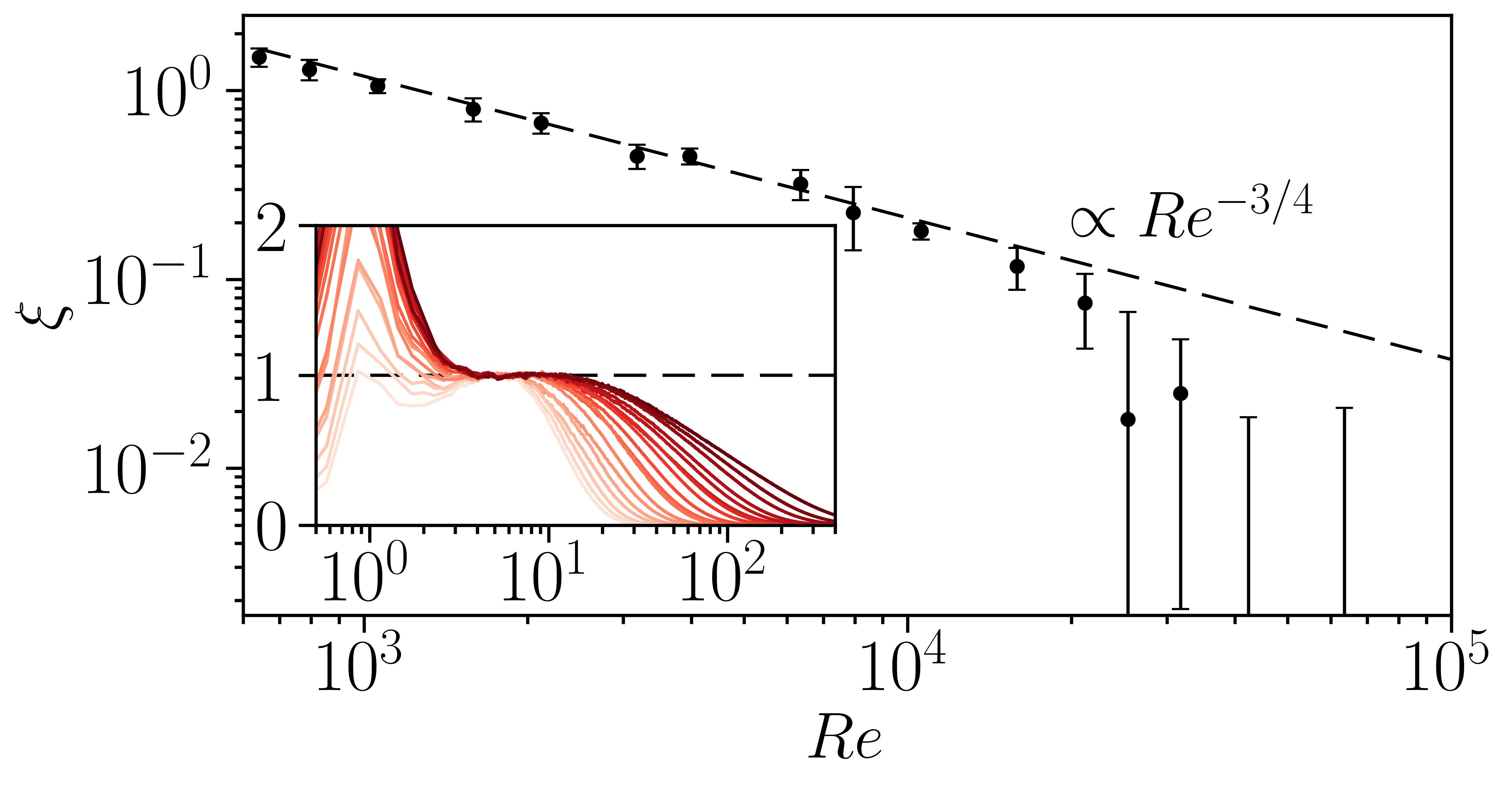}
\caption{Main plot: scaling exponent correction $\xi$ to the SPE spectrum
as a functions of $\text{Re}$. Inset: $E(k)$ compensated with $E_K(k)$ given by (\ref{eq12}). 
The runs follow the same color coding as in Fig.~\ref{fig2}.}
\label{fig4}
\end{figure}  

In Fig.~\ref{fig4}, we plot the dependence of $\xi$ on $\text{Re}$, together
with the spectra of Fig.~\ref{fig3} compensated with the expression
(\ref{eq12}). 
It is evident that, while for $\text{Re} \lesssim 10^4$, the exponent correction 
$\xi$ depends on $\text{Re}$ (approximately as $\text{Re}^{-3/4}$), for larger values 
of $\text{Re}$, the correction decreases much faster and becomes smaller than 
$5\%$ for $\text{Re}>2 \times 10^4$. 
In this limit, we also observe the convergence of 
the dimensionless constant to $C_K = 5.05 \pm 0.11$ (see Fig.~\ref{fig3}).

We remark that this behavior, which suggests the existence of a minimum
Reynolds number for the recovery of dimensional scaling is very different
to what was observed in the direct cascade of 3D NS turbulence, where 
Kolmogorov scaling is observed as soon as the spectrum displays a 
power-law behavior. 

\subsection{Predictability of the direct cascade}
\label{sec3b}

In this Section, we investigate the predictability of the cascade by
computing how two solutions $\theta(\bm{x},t)$ 
and $\theta'(\bm{x},t)$ separate in time on average. 
We will consider infinitesimally close solutions so that 
the average separation rate is given by the maximal Lyapunov exponent of the flow. 

Starting from a solution $\theta(\bm{x},t)$ of (\ref{eq1}) in a 
statistically stationary state, we generate a perturbed solution as
$\theta'(\bm{x},t)=\theta(\bm{x},t)+2\sqrt{\Delta} W({\bm x})$
where $W({\bm x})$ is a Gaussian random white noise with zero mean and unit
variance while $\Delta$ is a small parameter. 
The SPE error $E_{\Delta}$ is defined, for any time, as 
\begin{equation}
E_{\Delta}(t)=\frac{1}{2} \langle \delta \theta({\bm x},t) ^2 \rangle
\label{eq13}
\end{equation}
where the difference field is 
$\delta \theta=(\theta'-\theta)/\sqrt{2}$ and the normalization coefficient
$1/\sqrt{2}$ ensures that $E_{\Delta}=E$ for two completely 
uncorrelated fields. 
At initial time, by definition, we have $E_{\Delta}(t)=\Delta$. 
We measure the finite-time Lyapunov exponent by computing
the growth rate of the error
\begin{equation}
\gamma_{\tau}(t)=\frac{1}{2\tau} \ln 
\left(\frac{E_\Delta(t+\tau)}{\Delta} \right)
\label{eq14}
\end{equation}
and then by rescaling the perturbed field to the initial SPE error
\begin{equation}
\theta' \leftarrow \theta - \sqrt{\frac{\Delta}{E_\Delta}} 
\left(\theta - \theta'\right) \, .
\label{eq15}
\end{equation}
By repeating the steps (\ref{eq14}) and (\ref{eq15}) over many time
intervals of the same length $\tau$, we obtain a distribution of FTLE along 
the trajectory. The rescaling procedure (\ref{eq15}) ensures the
permanence of the perturbation in the exponential growth regime
when $\Delta$ and $\tau$ are sufficiently small \citep{vulpiani2009chaos}

From the definition (\ref{eq14}) one can compute the FTLE for
any time multiple of $\tau$, $T=n \tau$, simply by averaging
\begin{equation}
\gamma_{T}(t) = \frac{1}{n} \sum_{k=1}^{n} \gamma_{\tau}(t+k \tau)
\label{eq16}
\end{equation}
and the Lyapunov exponent is given by the average of FTLE over a very
long trajectory (and become independent of the initial condition)
\begin{equation}
\lambda = \lim_{T \to \infty} \gamma_{T}(t) \, .
\label{eq17}
\end{equation}

In general, the distribution of FTLE around the Lyapunov exponent, for 
sufficiently large $T$, follows the large deviation principle \citep{vulpiani2014large}
which states that
\begin{equation}
\rho(\gamma_T) = \frac{1}{N_T} e^{-T C(\gamma_T) }
\label{eq18}
\end{equation}
where $N_T$ is a normalizing factor and $C(\gamma_T)$ is the Cram\'er
function, independent of $T$ which, in general, 
vanishes at $\gamma_T=\lambda$ and is positive for $\gamma_T \neq \lambda$
\citep{boffetta2002predictability}.
For not too large fluctuations, the Cram\'er function can be approximated
by a quadratic form 
\begin{equation}
C(\gamma_T)\approx\frac{(\gamma_T-\lambda)^2}{2\Omega} \ .\
\label{eq19}
\end{equation}
where $\Omega$, proportional to the variance of the distribution 
$\rho(\gamma_T)$, is obtained from 
\begin{equation}
\Omega = \lim_{T \to \infty}T\langle(\gamma_T-\lambda)^2\rangle_{\gamma}
\label{eq20}
\end{equation}
where the brackets represents average is over the distribution (\ref{eq18}) while $\Omega$ is expected to be $T$-independent in the limit of large $T$.

\begin{figure}[!ht]
\centering\includegraphics[width=0.9\linewidth]{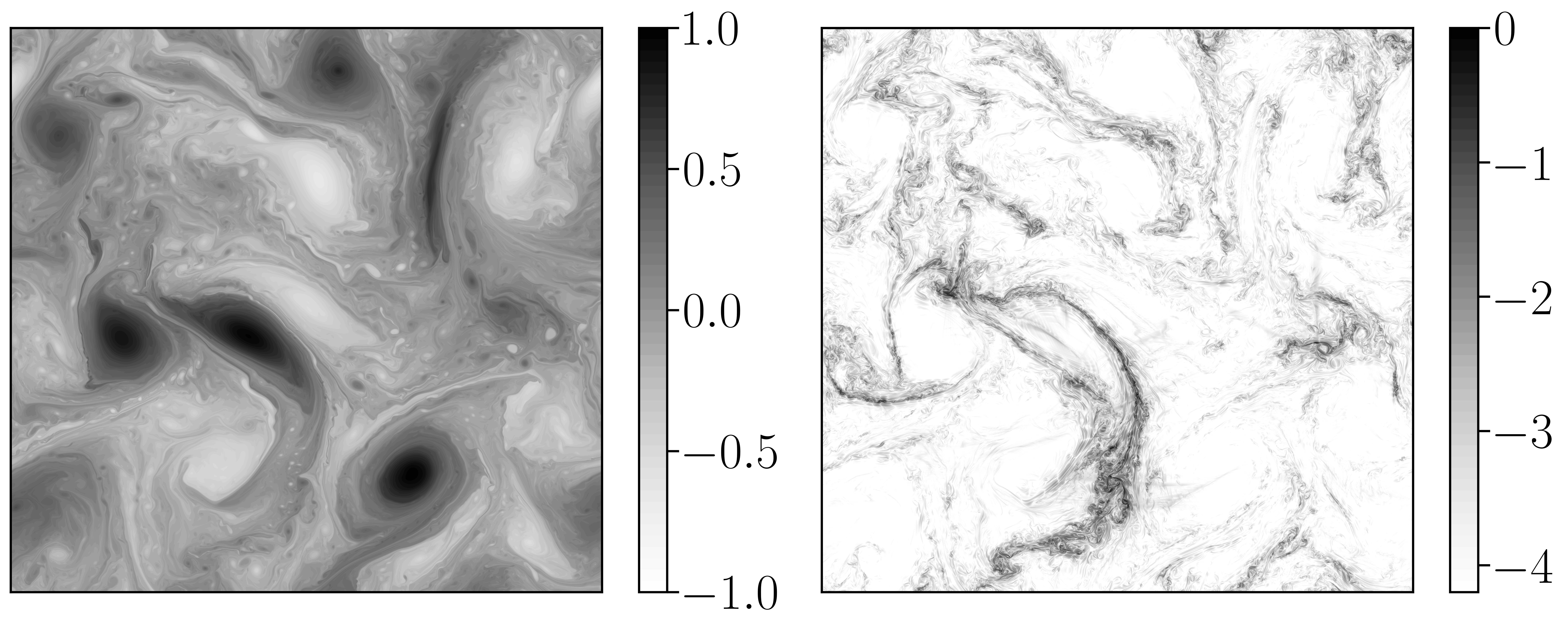}
\caption{Fields $\theta(\bm x)$ and $\delta\theta(\bm x)$ of run $D_3$ are shown in the left and right panels, respectively. The color code of the left panel is based on $\theta(\bm x)/\sup_{\bm x}|\theta(\bm x)|$. The right panel's color code is based on a log scale $(=\log_{10}(|\delta\theta(\bm x)|/\sup_{\bm x}|\delta\theta(\bm x)|))$ to facilitate visualization.}
\label{fig5}
\end{figure}  
We computed the FTLE for simulations at different Reynolds numbers corresponding to runs $B_2$, $B_4$, $C_2$, $C_3$, $C_4$, $C_5$, $D_1$, $D_2$, and $D_3$ of Table~\ref{tab1}. 
For all runs, we excluded from the statistics the initial transient during which the perturbation aligns with the most unstable direction of the system.
For the three cases at the highest Reynolds numbers, we compensated for the increased computational cost by averaging over nine independent, shorter simulations run in parallel, each of which with different realizations of the forcing $f(\bm{x})$ and initial perturbation noise $W(\bm{x})$. All rescaling times $\tau$ were kept around the Kolmogorov time scale of the simulation $\tau_\nu$, more precisely, between $\tau/\tau_\nu\in[0.5,0.8]$ depending on the run.

Figure~\ref{fig5} shows a representative realization of the field $\theta(\bm{x})$ along with its corresponding perturbation field $\delta\theta(\bm{x})$. The errors accumulate predominantly in filamentary zones between coherent structures. These regions are dominated by small-scale structures formed by energy transfer from larger scales \citep{pierrehumbert1994spectra}. Since such structures appear intermittently in time, the convergence of the FTLE statistics requires very long simulations, as discussed below. 

\begin{figure}[!ht]
\centering\includegraphics[width=0.75\linewidth]{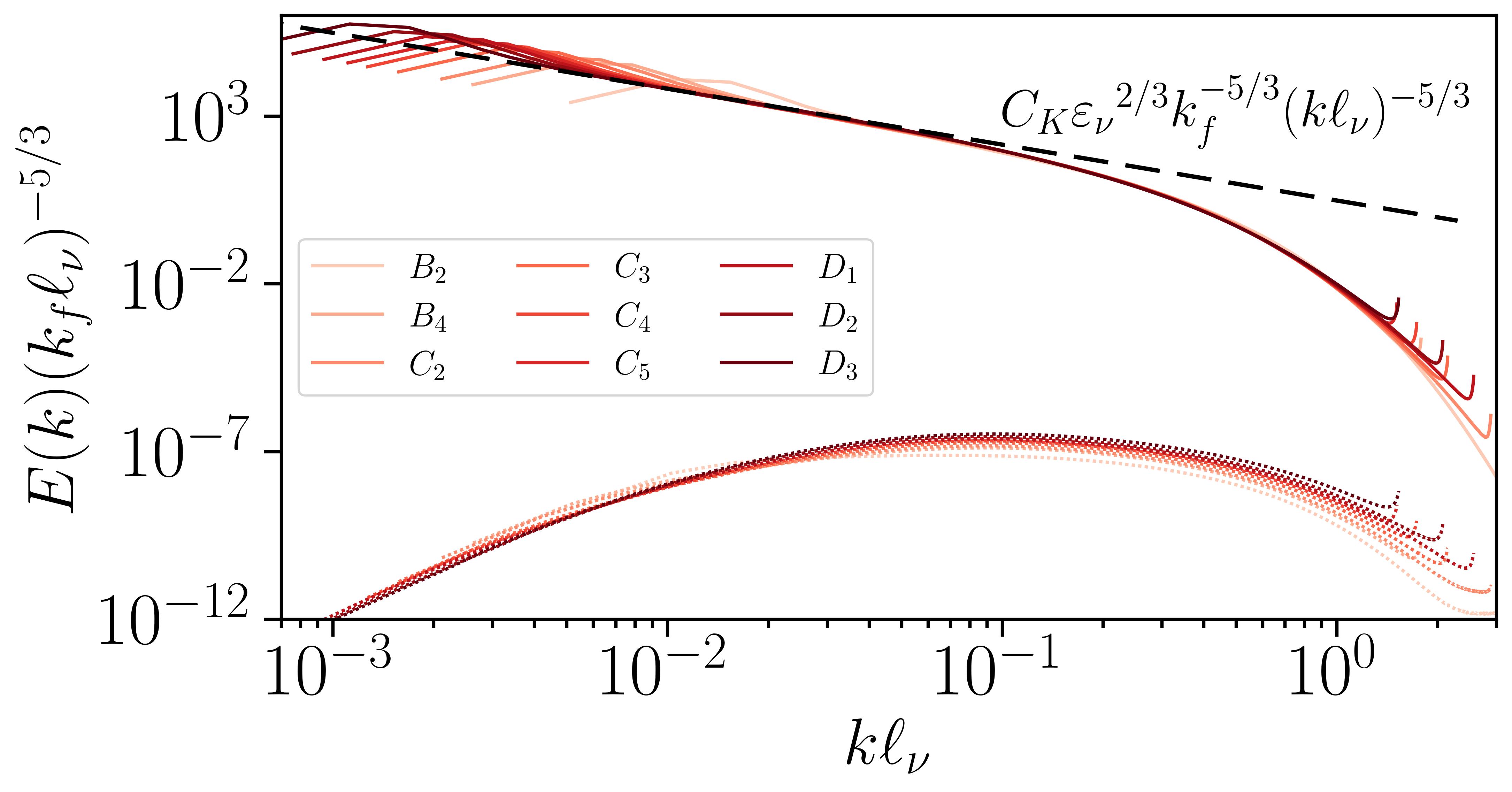}
\caption{Energy spectra $E(k)$ (full lines) and error spectra $E_\Delta(k)$ (dotted line) for the different runs as functions of $k\ell_\nu$. The error spectra are computed from the difference field $\delta \theta$ before the rescaling (\ref{eq15}) to the initial error.}
\label{fig6}
\end{figure}  

The distribution of the error field among scales, right before the rescaling, is shown in Fig.~\ref{fig6} where we plot, together with the energy spectra $E(k)$, the error spectra defined in a similar way as 
$E_{\Delta}(k)=\langle |\delta \hat{\theta}({\bf k})|^2 \rangle/2$. We observe that for all simulations, the error is concentrated at small scales $k \ell_{\nu} \simeq 0.1$ close to the dissipative range. We remark that the relative magnitude $E_{\Delta}(k)/E(k)$ is very small, ensuring that the perturbation remains in the linear regime.

\begin{figure}[!ht]
\centering\includegraphics[width=0.75\linewidth]{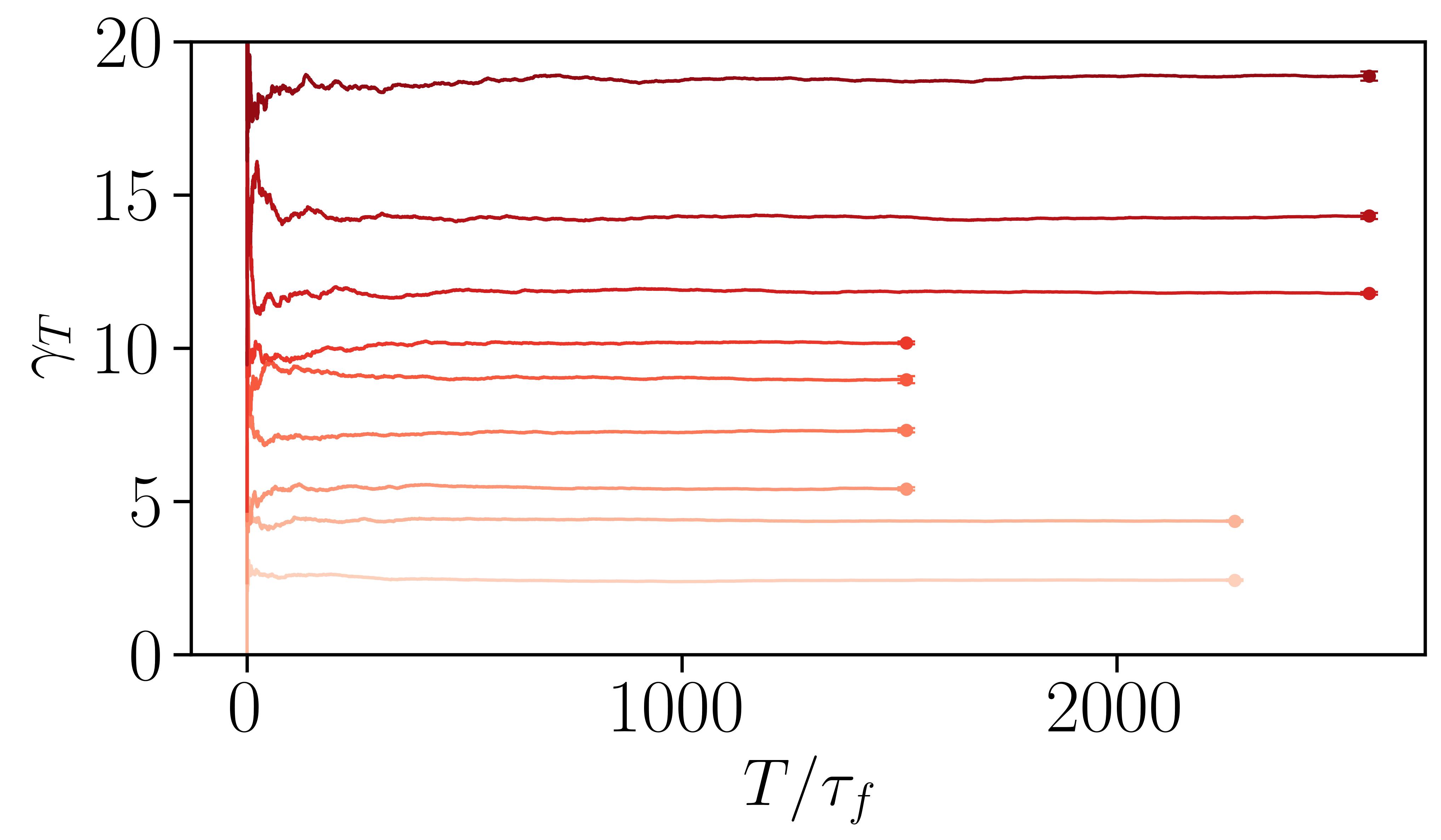}
\caption{Convergence of the FTLE as a function of the average
time. Light to dark colors represents runs at increasing Reynolds numbers $B_2$, $B_4$, $C_2$, $C_3$, 
$C_4$, $C_5$, $D_1$, $D_2$ and $D_3$.
}
\label{fig7}
\end{figure}

Figure~\ref{fig7} presents the FTLE for the different runs as a 
function of the average time $T$. 
We see that in all the cases, the average FTLE converges, after a long 
transient and for $T \gtrsim 200\tau_f$, to the asymptotic value,
which represents the Lyapunov exponent of the flow. 

From Fig.~\ref{fig7}, it is evident that the Lyapunov exponent increases 
with the Reynolds number. 
As discussed in Section~\ref{sec2}, we expect that $\lambda$ follows the Ruelle scaling \eqref{eq21}. 
The latter relies on the spectrum having a Kolmogorov-like power-law behavior. As shown in Fig.~\ref{fig4}, the scaling exponent $5/3$ is recovered only for $\text{Re} > 2 \times 10^4$. Ruelle's prediction is therefore expected to hold only for large $\text{Re}$.

\begin{figure}[!ht]
\centering\includegraphics[width=0.49\linewidth]{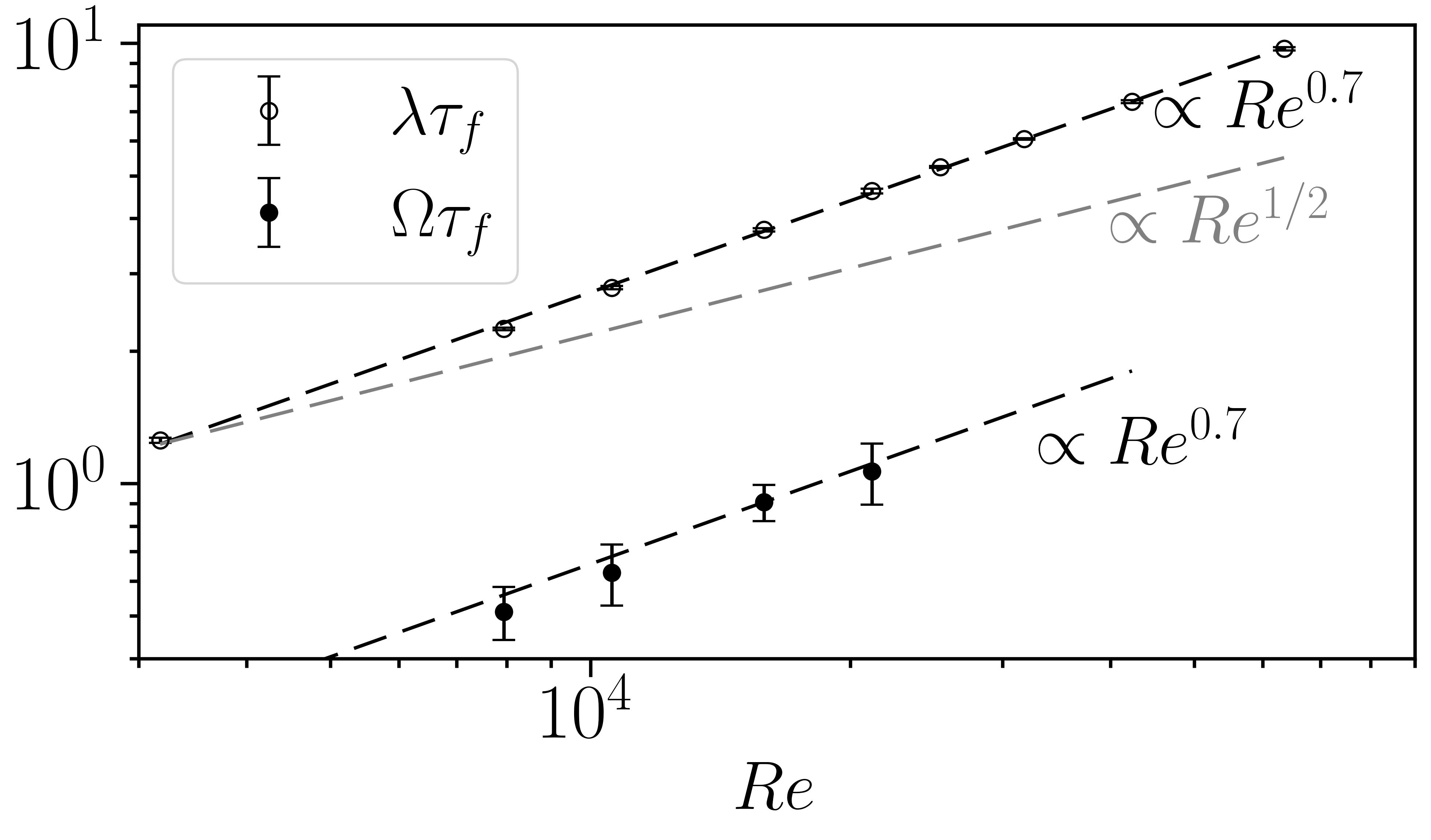}
\centering\includegraphics[width=0.49\linewidth]{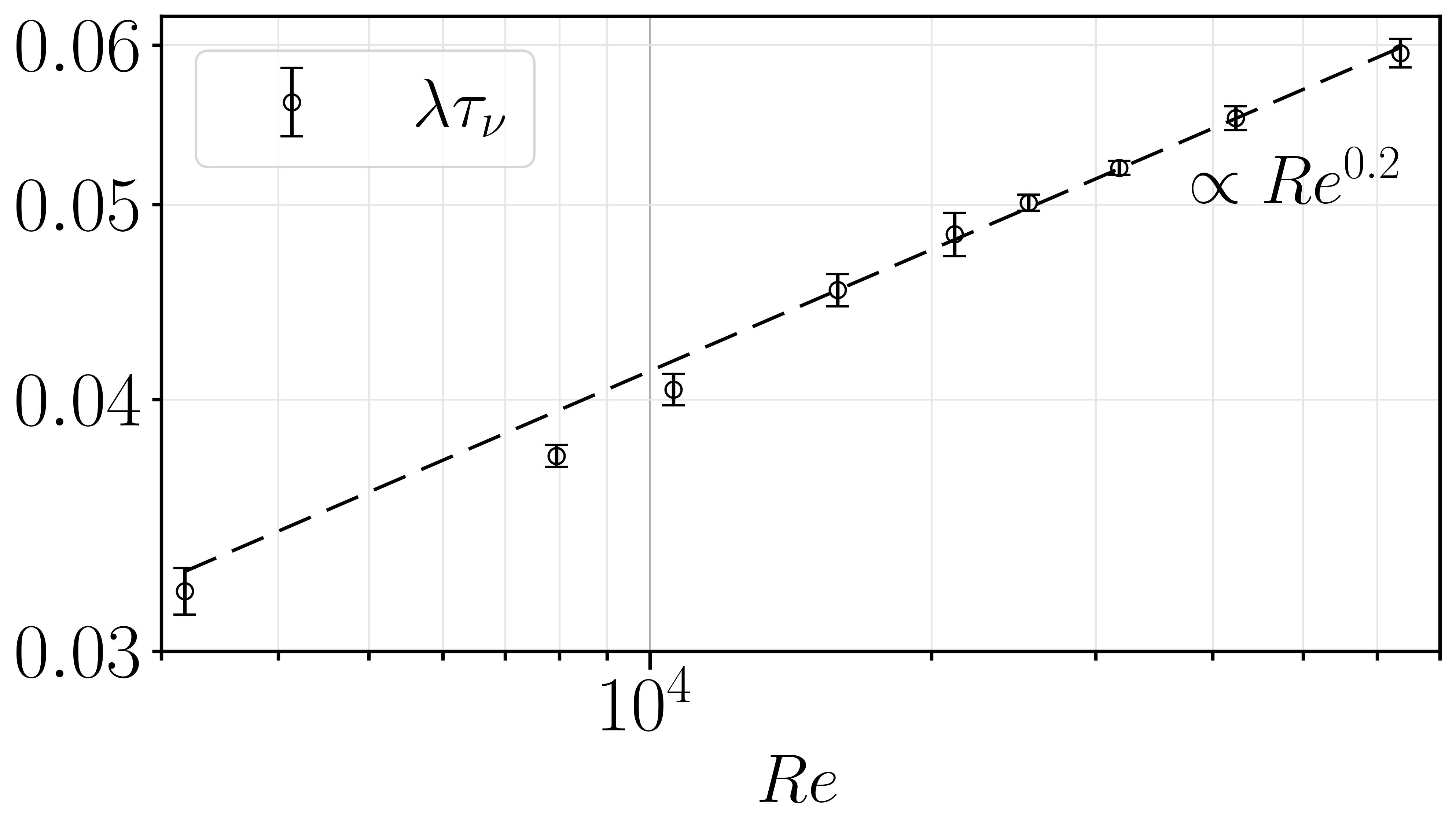}
\caption{Reynolds scaling of the mean FTLE $\mean{\gamma_T}=\lambda$. On the
left panel, nondimensionalization is made with $\tau_f$, while the right panel
uses $\tau_\nu$.}
\label{fig8}
\end{figure}

In Fig.~\ref{fig8}, we plot the Lyapunov exponents of our simulations
as functions of $\text{Re}$. We find that $\lambda$ grows with $\text{Re}$ faster 
than what predicted by (\ref{eq21}) and the best fit gives
$\lambda \tau_f \simeq \text{Re}^{0.7}$ or, equivalently, 
$\lambda \tau_{\nu} \simeq \text{Re}^{0.2}$.
Remarkably, the scaling persists also at $\text{Re}<10^4$, where the spectrum displays a significant correction to the Kolmogorov exponent $5/3$ (see Fig.~\ref{fig4}).

In Fig.~\ref{fig8} we also plot the scaled variance $\Omega$ as a 
function of $\text{Re}$, which displays a scaling law compatible with that of 
the Lyapunov exponent $\Omega \tau_f \simeq \text{Re}^{0.7}$. This 
indicates that, in the range of $\text{Re}$ investigated here, the ratio
$\Omega/\lambda$ is approximately constant ($0.24\pm 0.02$) and that the central
part of the Cram\'er function has a self-similar evolution with $\text{Re}$. 

A qualitatively similar behavior has been observed for the Lyapunov 
exponent of 3D turbulence 
\citep{boffetta2017chaos,mohan2017scaling,berera2018chaotic,ge2023production} 
with a correction to the dimensional scaling (\ref{eq21}) 
slightly smaller than in the present case,
$\lambda \propto Re^{0.64}$. 
We remark that the origin of this correction in 3D turbulence is still unclear since it cannot be simply attributed to intermittency. Indeed, the multifractal extension of the dimensional prediction  (\ref{eq21}) predicts, for 3D turbulence, an exponent smaller than $1/2$ \citep{aurell1996growth}.

\begin{figure}[!ht]
\centering\includegraphics[width=0.99\linewidth]{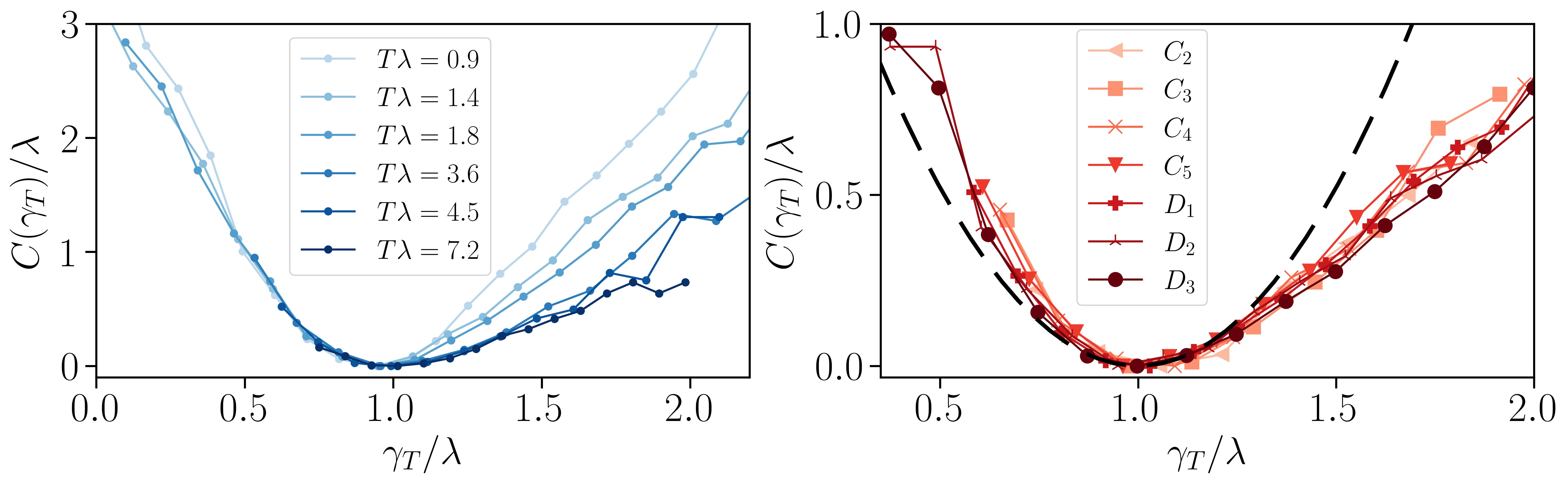}
\caption{Left: Cram\'er function for run $C_4$ and different times. Right: Cram\'er functions for different $\text{Re}$ at fixed $T\lambda=7$, dashed line represents the quadratic form $(x-1)^2/(2 \Omega/\lambda)$ with $\Omega/\lambda=0.24$.}
\label{fig9}
\end{figure}

In Figure~\ref{fig9} (left panel), we show the Cram\'er function $C(\gamma_{T})$ obtained from the distribution of FTLE~\eqref{eq18} as $C(\gamma_{T}) = - \frac{1}{T} \log(\rho(\gamma_T)/\rho(\lambda))$ at different times $T$ at $\text{Re} = 21200$ (run $C_4$). 
We notice that the convergence towards the asymptotic behavior of the left tail ($\gamma_T < \lambda$) is much faster than that of the right tail ($\gamma_T > \lambda$). 
At $T \gtrsim 2/\lambda$, the left tail is already converged, while the right tail achieves convergence only for $T \gtrsim 7/\lambda$.
The right panel of Figure~\ref{fig9} shows the comparison of the asymptotic behavior of the Cram\'er functions $C(\gamma_{T})$ computed at $T\lambda = 7$ for different $\text{Re}$ in the range $[10600, 63500]$.

Since $\lambda$ and $\Omega$ scale with $\text{Re}$ with the same exponent, we rescale both $\gamma_T$ and $C(\gamma_T)$ by $\lambda$ which, according to the quadratic approximation (\ref{eq19}), predicts the collapse to the function $(x-1)^2/(2 \Omega/\lambda)$.

Figure~\ref{fig9} shows that the quadratic approximation (\ref{eq19}) with $\Omega/\lambda = 0.24$ describes well the core of $C(\gamma_{T})$ (for $\gamma \simeq \lambda$), while we observe significant deviations for $\gamma_{T}>\lambda$. This means that the situations in which the dynamics of the system is strongly unpredictable (i.e. with $\gamma_{T} \gg \lambda$) are more frequent than what is expected by Gaussian statistics. Nonetheless, it is remarkable that the probability of these extreme deviations seem to be independent of the Reynolds numbers, as shown by the collapse of the right tail of $C(\gamma_{T})$ for different $\text{Re}$.

\section{Conclusions}\label{sec4}

In this study, we investigated the statistical properties and predictability of turbulence in the Surface Quasi-Geostrophic (SQG) model using high-resolution direct numerical simulations across a wide range of Reynolds numbers. Our analysis focused on two central aspects: the scaling behavior of the energy spectrum in the direct cascade of surface potential energy (SPE) and the chaotic dynamics characterized by finite-time Lyapunov exponents (FTLEs).

Our results indicate that, for $\text{Re}\gtrsim 2\times10^4$, the energy spectrum approaches the Kolmogorov-like scaling $E(k) \propto k^{-5/3}$. This observation, together with the convergence in Reynolds of the prefactor $C_K$ suggests that SQG turbulence exhibits a well-defined inertial range—similar to that of 3D Navier-Stokes turbulence, but it does so only at very large Reynolds numbers. This regime was not observed by earlier studies at moderate Reynolds numbers.

In terms of predictability, we showed that the Lyapunov exponent scales anomalously with the Reynolds number as $\lambda \propto \text{Re}^{0.7}$, exceeding the dimensional prediction $\lambda \propto \text{Re}^{1/2}$.  
The presence of an anomalous scaling is reminiscent of similar behavior observed in 3D Navier-Stokes turbulence, 
\citep{boffetta2017chaos,mohan2017scaling,berera2018chaotic,ge2023production}. 
In particular, it has been observed that $\lambda \propto \text{Re}^{0.64}$ \citep{boffetta2017chaos}. 
This is in contrast with the scaling of the Lyapunov exponent of Lagrangian trajectories $\lambda_L$, 
which has been found to be in agreement with the Ruelle's dimensional prediction 
$\lambda_L \propto \tau_\nu$ \citep{bec2006lyapunov} up to minor corrections 
which can be estimated from multifractal arguments.

The difference between Eulerian and Lagrangian predictability in NS turbulence can be 
ascribed to the fact that the Lagrangian Lyapunov exponent is determined by the viscous time-scale $\tau_\nu$, while the Eulerian predictability can be influenced also by the sweeping time-scale $\tau_E = \ell_\nu/U$, where $U$ is the RMS velocity \citep{ge2023production}. 
Considering that $\tau_E \ll \tau_\nu$ and that $\tau_\nu/\tau_E \simeq \text{Re}^{1/4}$, 
the dependence of $\lambda$ on $\text{Re}$ can be rewritten as a combination of two time scales 
$\lambda = \tau_\nu^{-1} \text{Re}^\xi = \tau_\nu^{-1+4\xi}\tau_E^{-4\xi}$. 
The Ruelle scaling is recovered for $\xi=0$, while for 
$\xi=1/4$ the predictability is determined by the sweeping time. 
Our results in SQG give $\xi^{SQG}=0.2$ 
which is larger than the value observed in 3D NS $\xi^{NS} = 0.14$, 
suggesting that the sweeping effect has a stronger influence 
on the Eulerian predictability in SQG than in NS.

It is worth to note that the dimensional arguments discussed above do not consider the effects of intermittency corrections.  It would be interesting to develop a multifractal-like approach for the direct cascade of SQG to check how the observed intermittency \citep{valade2025surface} affects the dimensional scaling of the Lyapunov exponent.

Beyond the average growth rate of infinitesimal perturbations, we investigated the statistical properties of FTLEs. Notably, both the variance and the shape of the associated Cram\'er function exhibit self-similar behavior across Reynolds numbers when appropriately rescaled. The ratio $\Omega / \lambda$ remains approximately constant across $\text{Re}$, indicating a form of universality in the core of the FTLE distribution.

Although we have studied the predictability problem from the point of view of the exponential growth of infinitesimal perturbations, it would be very interesting to investigate in detail the complementary regime of large errors and the statistics of finite-size Lyapunov exponents \citep{boffetta2017chaos}. Recent results have been obtained in the case of a decaying SPE cascade in SQG \citep{valade2024anomalous}, where the authors were able to connect the hyperdiffusive behavior of Lagrangian fluid parcels with the anomalous diffusion of the system.

\section*{Acknowledgement}

We acknowledge HPC CINECA for computing resources within the INFN-CINECA Grants INFN24-FieldTurb and INFN25-FieldTurb.

\section*{Funding} 

This work was supported by Italian Research Center on High Performance Computing Big Data and Quantum Computing (ICSC), project funded by European Union - NextGenerationEU - and National Recovery and Resilience Plan (NRRP) - Mission 4 Component 2 within the activities of Spoke 3 (Astrophysics and Cosmos Observations). 

\section*{Declaration of interests} 

The authors report no conflict of interest.

\section*{Data availability statement} 

The data that support the findings of this study are available upon request.

\section*{Author ORCID} 

V.J. Valad\~{a}o, https://orcid.org/0000-0002-7603-5969; F. De Lillo, https://orcid.org/0000-0002-1327-695X; S. Musacchio, https://orcid.org/0000-0002-4564-8527; G. Boffetta, \\ https://orcid.org/0000-0002-2534-7751.

\section*{Author contributions} 

V.J.V. performed the simulations and analyzed the data. All authors contributed equally in the discussions to reach conclusions and to write the paper.

\bibliographystyle{jfm}
\bibliography{biblio}

\end{document}